\documentclass[journal=jpcl,manuscript=letter]{achemso}
\usepackage{amsfonts}
\usepackage{amssymb}
\usepackage{bbold}
\usepackage{graphicx}
\usepackage{amsmath}
\usepackage{tikz}
\usepackage{adjustbox}
\usepackage[normalem]{ulem}
\usetikzlibrary{shapes.geometric, arrows}
\usepackage[english]{babel}
\usepackage{color}
\usepackage[version=3]{mhchem} 
\usepackage{url}
\usepackage[colorlinks,citecolor=black,urlcolor=blue,linkcolor=black]{hyperref} 
\usepackage{threeparttable}
\usepackage{makecell}
\usetikzlibrary{arrows,shapes,positioning,shadows,trees}

\author{Yang Zhao}
\affiliation{Hebei Key Laboratory of Photophysics Research and Application, Hebei Normal University, Shijiazhuang, Hebei 050024, China}
\email{zhaoyang22@hebtu.edu.cn}
\author{Lipeng Chen}
\affiliation{Zhejiang Laboratory, Hangzhou 311100, China}
\email{chenlp@zhejianglab.org}

\title{Polarization dynamics of the spin-boson model in the shifted boson Hilbert space}

\begin{document}

\begin{tocentry}
\centering
\includegraphics[scale=0.19]{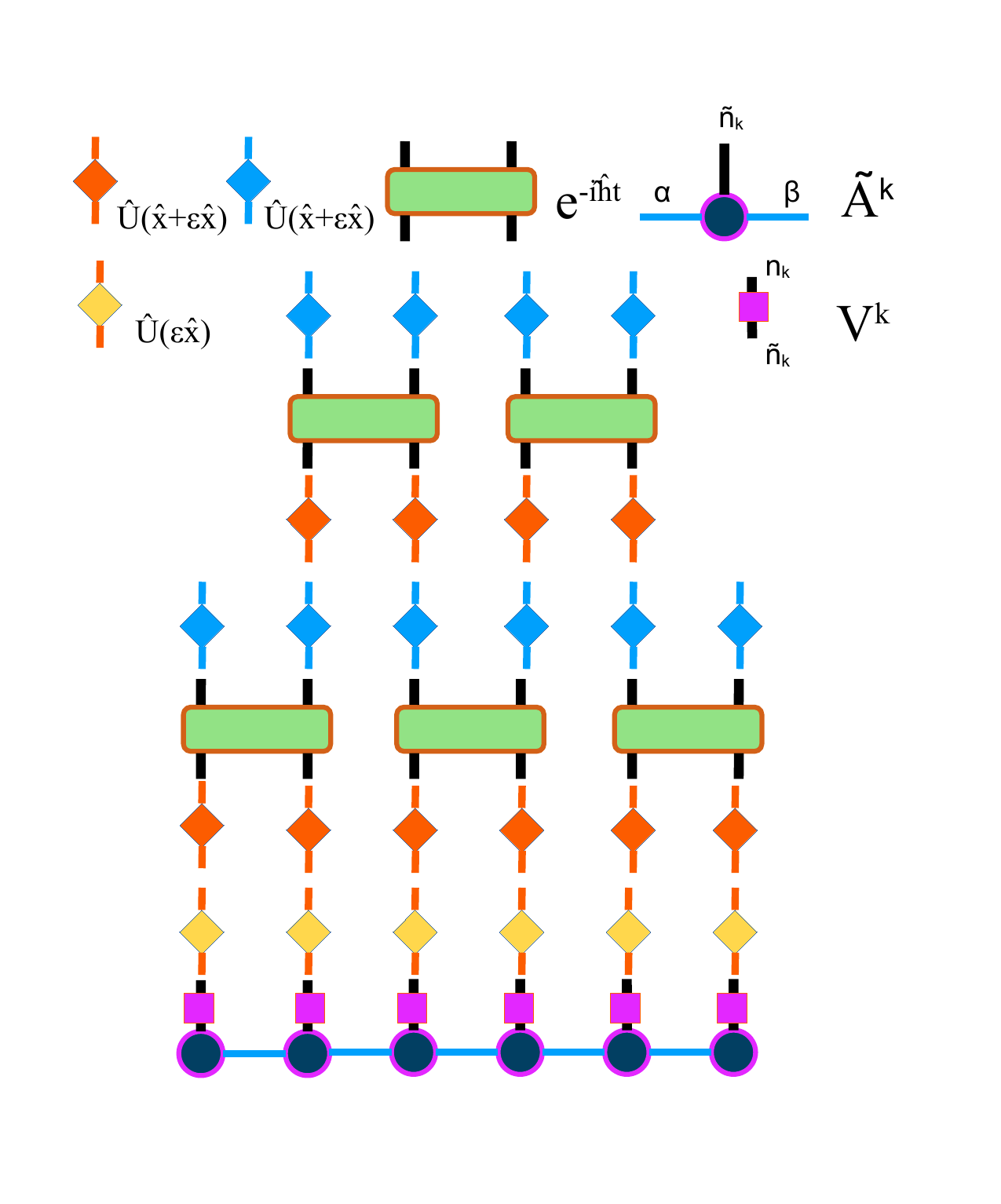}
\end{tocentry}

\begin{abstract}
Faithfully simulating the dynamics of open quantum systems requires efficiently addressing the challenge of an infinite Hilbert space. Inspired by the shifted boson operator technique used in ground-state studies of the spin-boson model (SBM), we develop a novel algorithm that integrates a shifted optimized boson basis with the time-evolving block decimation method. We validate our approach by accurately reproducing the polarization dynamics of the sub-Ohmic SBM at a significantly reduced computational cost. For the Ohmic SBM, we demonstrate that the time-evolved final state converges precisely to the variational prediction at both zero and finite temperatures. Furthermore, our method reveals a new aperiodic pseudocoherent phase in the super-Ohmic SBM with an initially polarized bath. This work establishes an efficient and powerful approach for simulating the real-time dynamics of open quantum systems. 
\end{abstract} 

Understanding irreversible quantum phenomena, such as relaxation and decoherence, represents a central challenge in the theory of open quantum systems. \cite{LeggettRMP,WeissBook} A paradigmatic model for studying these environment-induced effects is the SBM, which describes a two-level system (TLS) coupled to a bosonic heat bath. The environment's influence is encoded in its spectral density, $J(\omega)\sim\omega^s$, where the exponent $s$ categorizes the bath as sub-Ohmic ($0<s<1$), Ohmic ($s=1$), or super-Ohmic ($s>1$). 

Accurately simulating the SBM dynamics is notoriously difficult due to the bath's intrinsic non-perturbative and non-Markovian character. Numerically exact methods developed to meet this challenge fall into two broad categories. The first group uses a reduced system description, capturing bath effects through correlation functions; this includes the hierarchical equations of motion (HEOM),\cite{Tanimura2006,Tanimura2020} the quasi-adiabatic propagator path integral (QUAPI),\cite{QUAPI1995,QUAPI2017,QUAPI2023} and real-time path integral Monte Carlo (PIMC).\cite{PIMC2013} The second group comprises wave function methods that explicitly discretize the bath degrees of freedom to simulate the full system-bath dynamics. Powerful examples include the (multilayer) multiconfiguration time-dependent Hartree method ((ML-)MCTDH),\cite{MCTDH2000,MLMCTDH2003,MLMCTDH2006,MLMCTDH2008} the time-dependent density matrix renormalization group (t-DMRG),\cite{TDDMRG2010} the time-evolving block decimation (TEBD) algorithm,\cite{TEBD,TEBD2} the multiple Davydov ansatz,\cite{Davydov2022,Davydov2023} and various matrix product state (MPS) techniques.\cite{MPS2013,MPS2016,MPS2019}

Despite its simplicity, the SBM exhibits rich and nontrivial dynamics. For the Ohmic case ($s=1$), increasing the system-bath coupling strength induces a well-known transition from coherent oscillations to incoherent relaxation, culminating in a quantum phase transition.\cite{LovettNC2018,WHBJPCA2019} The sub-Ohmic regime ($0<s<1$) presents a greater challenge, as strong coupling to low-frequency modes enhances non-Markovian effects. Here, numerical studies show a similar progression from damped oscillation to incoherence and then to a localized phase,\cite{WHBNJP2008,WHBCP2010} though the dynamics are also profoundly sensitive to the initial bath preparation.\cite{PRLdp0,MPS2016,ThorwartPRB2010,ZhaoYangJCP2016,ThorwartPRL2022,LipengChenJCP2023} For instance, with an initially polarized bath, nonequilibrium coherent dynamics can persist for any coupling strength if $s<1/2$.\cite{PRLdp0,PIMCPRB2013} This dependence on initial conditions has been systematically mapped, revealing dynamic phase diagrams containing coherent, pseudo-coherent, and incoherent phases.\cite{ThorwartPRL2022,LipengChenJCP2023} In super-Ohmic baths ($s>1$), one observes pseudo-coherent dynamics at strong coupling, with a non-monotonic dependence on the exponent $s$.\cite{NalbachPRA2023}

Obtaining reliable dynamical properties of the SBM requires an adequate truncation of the bosonic Hilbert space. While Guo \textit{et al}. \cite{GuoPRL2012} utilized a variational MPS approach with a shifted optimized boson basis (OBB) to accurately map the ground-state phase diagram of the one and two-bath SBM, their work focused on static properties. To address dynamics, we propose a new method that combines the time-dependent variational principle (TDVP) with the TEBD algorithm, integrated with the shifted OBB protocol. This approach naturally describes the polarized heat bath and yields highly accurate polarization dynamics at a significantly reduced computational cost. 

The Hamiltonian of the SBM is given by 
\begin{equation}\label{SBMH}
\hat{H}=-\frac{\Delta}{2}\hat{\sigma}_x-\frac{\epsilon}{2}\hat{\sigma}_z+\sum_p\omega_p\hat{b}_p^{\dagger}\hat{b}_p+\frac{\hat{\sigma}_z}{2}\sum_p\lambda_p(\hat{b}_p^{\dagger}+\hat{b}_p),
\end{equation}
where $\epsilon$ and $\Delta$ denote the energy bias and tunneling constant, respectively. The Pauli matrices are represented by $\hat{\sigma}_i(i=x,y,z)$, and $\hat{b}_p^{\dagger}$($\hat{b}_p$) is the creation (annihilation) operator for the $p$-th bath mode with frequency $\omega_p$. The strength of the system-bath coupling, $\lambda_p$, is characterized by the spectral density function 
\begin{equation}
J(\omega)=\pi\sum_p\lambda_p^2\delta(\omega-\omega_p)=2\pi\alpha\omega_c^{1-s}\omega^s\theta(\omega_c-\omega).
\end{equation}
Here, $\omega_c$ is the cutoff frequency, $\alpha$ is the dimensionless coupling strength, $s$ is the spectral exponent, $\theta(x)$ is the Heaviside step function. To adapt this model for the TEBD method, we employ an orthogonal polynomial mapping \cite{Lmaping} to transform the Hamiltonian into a semi-infinite one-dimensional chain
\begin{equation}\label{SBMHMap}
\hat{H}=-\frac{\Delta}{2}\hat{\sigma}_x-\frac{\epsilon}{2}\hat{\sigma}_z+\frac{\hat{\sigma}_z}{2}\eta_1(\hat{b}_1^{\dagger}+\hat{b}_1)+\sum_{k=1}^{L-2}\left[\omega_k\hat{b}_k^{\dagger}\hat{b}_k+t_k\left(\hat{b}_k^{\dagger}\hat{b}_{k+1}+\hat{b}_{k+1}^{\dagger}\hat{b}_k\right)\right].
\end{equation}
The effective system-bath coupling to the first chain site is  $\eta_1=\sqrt{\int_0^{\omega_c}\frac{J(\omega)}{\pi}d\omega}=\sqrt{\frac{2\alpha\omega_c^2}{1+s}}$. For a chain of length $L$, the on-site energies  $\omega_k$ and nearest-neighbor couplings $t_k$ are given by 
\begin{eqnarray}
  \omega_k &=& \frac{\omega_c}{2}(1+\frac{s^2}{(s+2k-2)(s+2k)}), \nonumber\\
  t_k &=& \frac{\omega_ck(s+k)}{(s+2k)(1+s+2k)}\sqrt{\frac{s+2k+1}{s+2k-1}}. 
\end{eqnarray}

For the 1D chain mapping Hamiltonian of Eq.~\eqref{SBMHMap}, one can introduce the local eigenstate $|n_k\rangle$ of the occupation number operator $\hat{n}_k = \hat{b}^\dagger_k \hat{b}_k$, which satisfies $\hat{n}_k |n_k\rangle = n_k |n_k\rangle$. To render calculations feasible, the infinite bosonic Hilbert space at each site $k$ is truncated by imposing an upper bound $d_k$ on the occupation number ($0 \leq n_k < d_k$). The MPS representation of the chain-mapping Hamiltonian's wavefunction is 
\begin{equation}
|\Psi\rangle=\sum_{\sigma=\uparrow,\downarrow}\sum_{\{\vec{n}\}}A^{0}[\sigma]A^{1}[n_1]\cdots{A}^{L-1}[n_{L-1}]|\sigma\rangle|\vec{n}\rangle,
\end{equation}
where $|\sigma\rangle=|\uparrow\rangle,|\downarrow\rangle$ are the eigenstates of $\hat{\sigma}_z$, and $|\vec{n}\rangle = |n_1, \cdots, n_{L-1}\rangle$ constitutes a basis of boson-number eigenstates. We then implement the OBB technique,\cite{GuoPRL2012} which compresses the local basis by representing the A-matrix elements as 
\begin{equation}
(A^{k}[n_k])_{\alpha\beta}=\sum_{\tilde{n}_k=0}^{d_{\mathrm{opt}}-1}(\tilde{A}^{k}[\tilde{n}_k])_{\alpha\beta}V^{k}_{\tilde{n}_kn_k}\quad\quad\quad(k\geq{1}).
\end{equation}
Here, the transformation matrix $V^{k}$ maps the local eigenstate $|n_k\rangle$ to the optimized basis $|\tilde{n}_k\rangle=\sum_{n_k=0}^{d_k-1}V_{\tilde{n}_kn_k}^{k}|n_k\rangle$, where $0\leq{\tilde{n}_k}<d_{\mathrm{opt}}$. This compression is highly effective for capturing the quantum critical behavior of the SBM \cite{GuoPRL2012} (see Supporting Information for technical details).    

Although the OBB protocol allows for a larger effective local basis, accurately describing the critical phenomena of the SBM requires an even larger number of states. This is achieved by incorporating explicit shifts of the oscillator coordinates. We shift the coordinate $\hat{x}_k = (\hat{b}^\dagger_k + \hat{b}_k)/\sqrt{2}$ on each site $k$ by its equilibrium expectation value $\langle\hat{x}_k\rangle$. This defines a new set of creation and annihilation operators: $\hat{b}'^\dagger_k = \hat{b}^\dagger_k + \langle \hat{x}_k \rangle/\sqrt{2}$ and $\hat{b}'_k = \hat{b}_k + \langle \hat{x}_k \rangle/\sqrt{2}$. The Hamiltonian in this shifted basis becomes $\hat{H}'(\hat{b}^\dagger_k, \hat{b}_k) = \hat{H}(\hat{b}'^\dagger_k, \hat{b}'_k)$. While Ref.~\citenum{GuoPRL2012} manually updates $\langle \hat{x}_k \rangle$ during the OBB sweeping procedure, we propose an alternative integration with the TEBD method. This approach offers a more systematic, albeit computationally more demanding, path to convergence (see details in the Supporting Information).    

\begin{figure}
  \centering
  \includegraphics[width=0.9\linewidth]{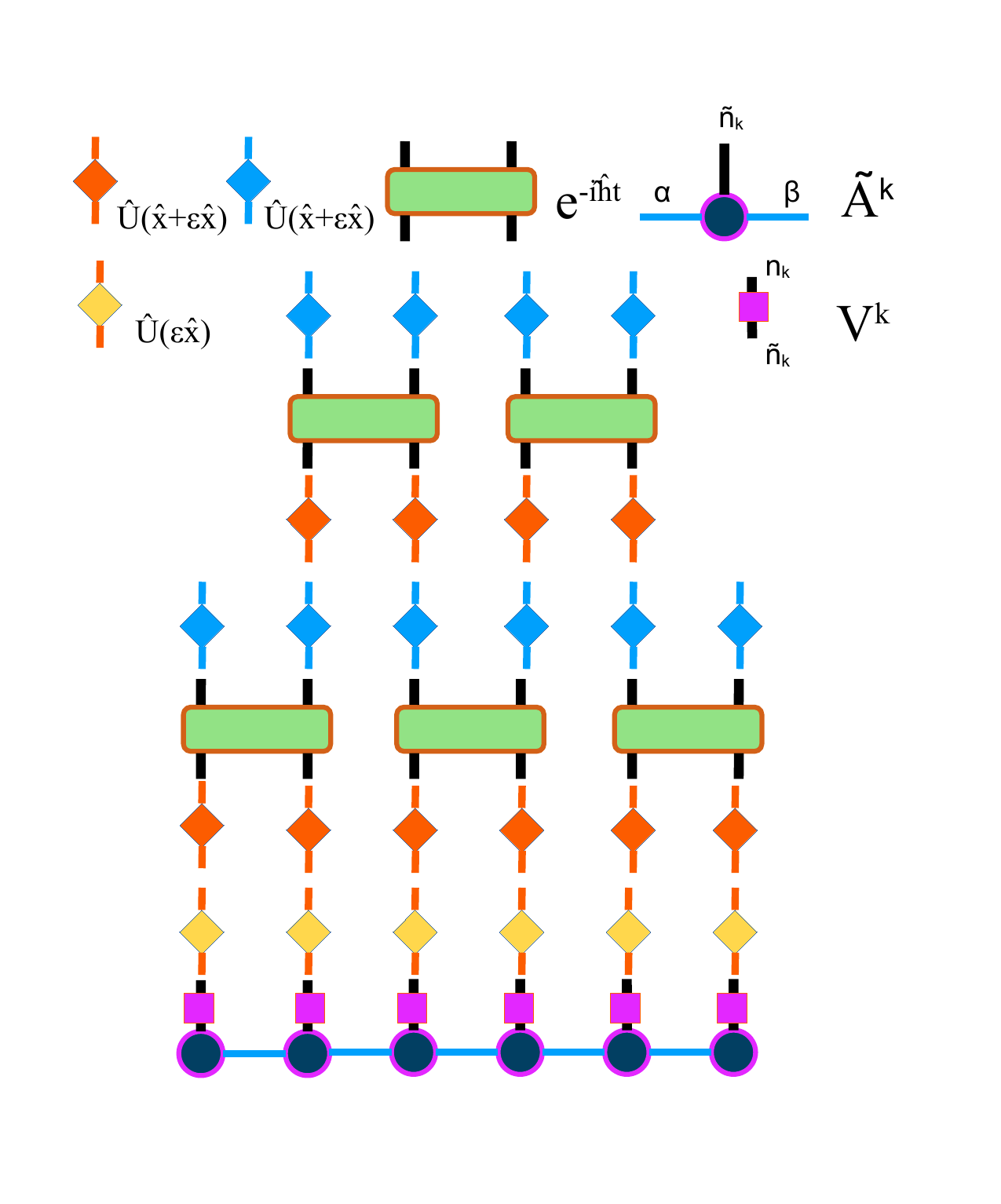}
  \caption{Tensor network diagram for the TEBD algorithm incorporating shifted bosonic operators. The time evolution operators (various shapes) and shifted boson operators are represented by distinct nodes. Pink squares denote the transformation matrices for the OBB, while dark purple dots represent the local tensors $(\tilde{A}^k[\tilde{n}_k]) _{\alpha\beta}$ of the MPS, with legs indicating their indices.}\label{TEBD}
\end{figure}

We now describe the incorporation of boson shifts into the TEBD algorithm. The procedure begins by using DMRG or TEBD to compute the ground state of the boson bath coupled to a fully polarized spin. This yields the initial wavefunction for time evolution, $|\tilde{\Psi}'(t=0)\rangle$, expressed in a shifted boson basis defined by the displacements $\langle \hat{x}_k \rangle$ and their corresponding transformation matrices $V^k_{\tilde{n}n}$. To perform time evolution within this shifted basis, the time evolution operator  $\exp(-i\hat{H}t)$ must be transformed accordingly. This is achieved by inserting the shift  operator $\hat U(\hat x_k)=\exp(\langle\hat x_k\rangle(\hat{b}^\dagger_k-\hat{b}_k)/\sqrt{2})$ at each site $k$ into the tensor network. First, $\exp(-i\hat{H}t)$ is decomposed via a Trotter-Suzuki expansion\cite{TSD} into a product of local two-site operators, $\exp(-i\hat{h}_{k,k+1}t)$, as depicted by the rectangles in Fig.~\ref{TEBD} (see details in the Supporting Information). The shift operators (represented by rhombuses in Fig.~\ref{TEBD}) are then applied to transform each local term into the shifted basis as $\hat{U}(\hat{x}_k) \otimes \hat{U}(\hat{x}_{k+1}) \exp(-i\hat{h}_{k,k+1}t)\hat{U}^\dagger(\hat{x}_{k+1}) \otimes \hat{U}^\dagger(\hat{x}_{k})$. Repeating this procedure for all the local time evolution operators completes the transformation of the full time-evolution operator $\exp(-i\hat Ht)$. 

To further increase the effective boson number, we apply an infinite shift operator. An operator $\prod_k \hat{U}(\varepsilon \hat{x}_k)$, where $\varepsilon$ is a small constant, is applied to the initial state, transforming it to $|\tilde{\Psi}(t=0)\rangle = \left[ \prod_k \hat{U}(\varepsilon \hat{x}_k) \right] |\tilde{\Psi}'(t=0)\rangle$. Consequently, the local shift operator becomes $\hat{U}(\hat{x}_k + \varepsilon\hat{x}_k)$, corresponding to a new displacement of $\langle (1+\varepsilon)\hat{x}_k \rangle / \sqrt{2}$. Since the boson number scales with $\langle \hat{x}_k \rangle^2$, this transformation increases the local occupation by a factor of $(1+\varepsilon)^2$, which significantly facilitates the subsequent time evolution. The resulting tensor network (Fig.~\ref{TEBD}) is then contracted layer-by-layer using the standard TEBD algorithm. The performance and accuracy of this approach depend on the choice of $\varepsilon$; we use a value of $\varepsilon = 0.1$ throughout this work. The associated numerical errors introduced by this choice are discussed in the Supporting Information.

\begin{figure}
  \centering
  \includegraphics[width=1.0\linewidth]{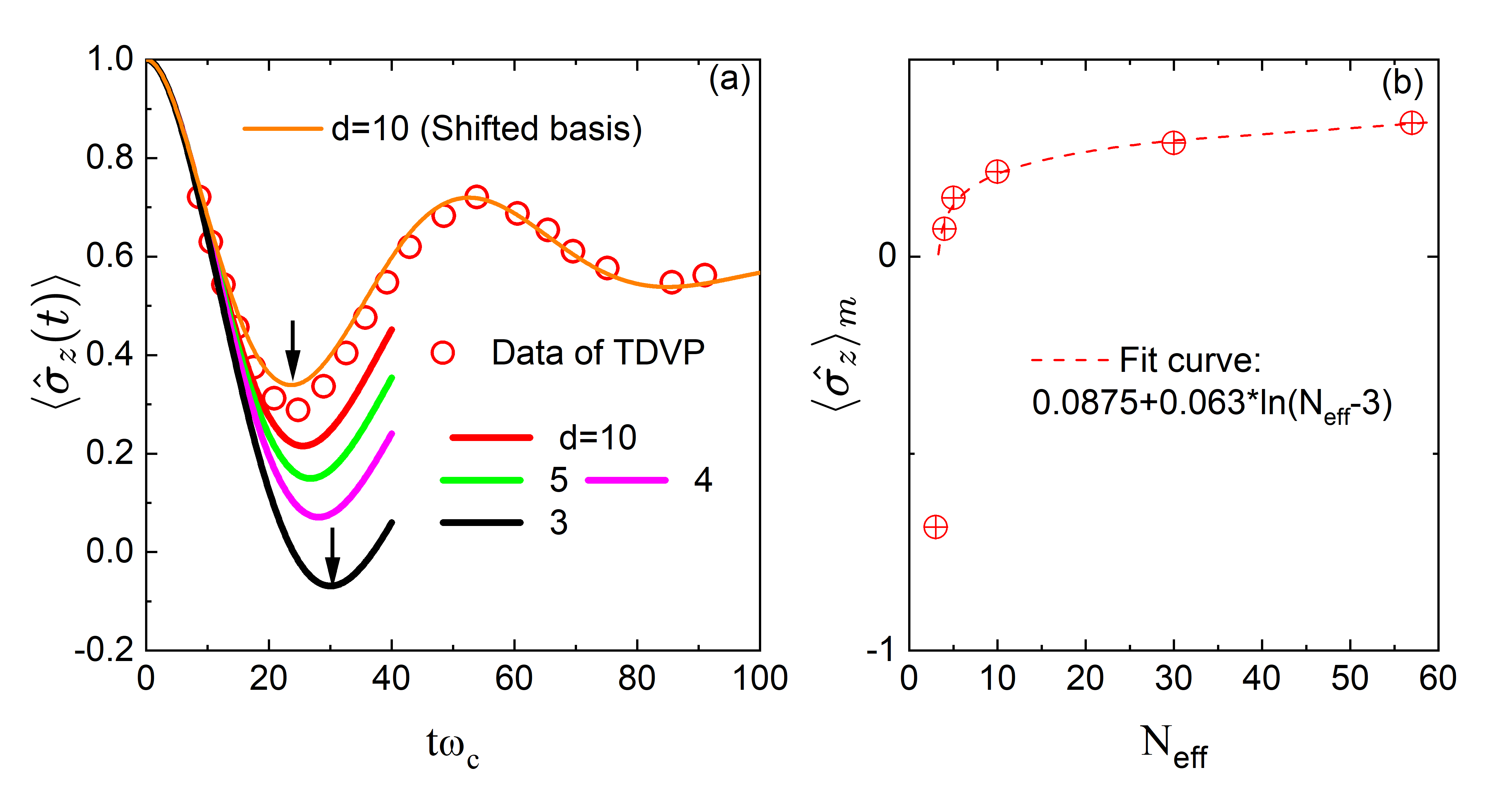}
  \caption{(a) Spin dynamics of the sub-Ohmic SBM at various $d$, red circles denote TDVP results extracted from Ref.~\citenum{Chin2016}. (b) The first local minimum of the curves in (a), $\langle\hat{\sigma}_z\rangle_m$, as a function of the effective boson number, $N_{\rm eff}$. The other parameters are $s=0.25$, $\alpha=0.03$, $L=30$.}\label{zero}
\end{figure}

We employ the TEBD method to compute the dynamics of the spin polarization, $\langle\hat \sigma_z(t)\rangle$, by evolving the initial state $|\tilde\Psi(0)\rangle$ under the effective Hamiltonian $\hat{H}^{'}$. To benchmark our approach in the shifted basis, we calculate $\langle\hat \sigma_z(t)\rangle$ for the same initial condition (a polarized boson bath) in both the shifted and unshifted Fock spaces. If not otherwise stated, we take $\Delta=0.1$, $\epsilon=0$, $\omega_c=1.0$, $\Delta t=0.1$ with a chain length of $L=30$ and matrix dimension of the MPS, $D_c=60-130$, $d=3-10$ ($d_{\rm opt}=d$) in all simulations.        

We begin by analyzing the case of zero temperature. Figure~\ref{zero} (a) presents $\langle\hat \sigma_z(t)\rangle$  for the sub-Ohmic SBM with $s=0.25$, $\alpha=0.03$ under various values of $d$, calculated in the unshifted boson basis. For conciseness, we show only the initial part of the curve. As expected, $\langle\hat \sigma_z(t)\rangle$ exhibits typical damped oscillations. The time of the first local minimum, $t_s$, shifts to earlier values as $d$ increases, a trend also visible in Figure~\ref{zero} (b). Furthermore, the value of this first minimum, $\langle\hat\sigma_z\rangle_m$, increases with $d$, indicating that it can serve as an indicator for the effective boson number. For a direct comparison with previous MPS results, we extracted the data calculated using $d=30$ from Ref.~\citenum{Chin2016} and plotted it in Fig.~\ref{zero} (a). $\langle\hat\sigma_z(t)\rangle$ calculated by the TEBD method converges toward this benchmark curve as $d$ increases. We then introduced shifts into the boson operators based on the unshifted results at $d=10$ and repeated the TEBD calculation. This approach produced results in good agreement with those from Ref.~\citenum{Chin2016}. Notably, using the shift boson operators with $d=10$ yielded a first minimum $\langle\hat\sigma_z\rangle_m$ that exceeded the benchmark value at $d=30$. This demonstrates that our method achieves comparable performance with a significantly smaller $d$ than algorithms using the unshifted boson basis.   

As noted in Ref.~\citenum{Thorwartdp}, the polarization dynamics of a spin in a polarized bath first evolves into a transient quasi-equilibrium state. In this state, $\langle\hat \sigma_z(t)\rangle$ oscillates around a finite value before undergoing a very slow decay to zero. The quasi-equilibrium state persists throughout the sub-Ohmic regime, even in the spin-localized phase. It is natural to expect that a greater number of boson modes would more easily drag the spin magnetic moment back to its initial value, which is consistent with the behavior shown in Fig.~\ref{zero} (a). To quantify this, we numerically estimate the effective local boson number, $N_{\rm eff}$, by fitting the growth of the first local minimum, $\langle\hat \sigma_z\rangle_m$, as indicated in Fig.~\ref{zero} (b). In the unshifted boson Fock space, $N_{\rm eff}$ is simply defined as the local dimension $d$. In the shifted boson space, however, $N_{\rm eff}$ must be obtained by fitting the scaling of $\langle\hat \sigma_z\rangle_m$ with respect to $d$. Generally, $\langle\hat \sigma_z\rangle_m$  is proportional to the system-bath coupling strength, which itself is proportional to the local boson occupation number and, consequently, the shift $\langle\hat{x}_k\rangle$. As shown in Fig.~\ref{zero} (b), the dependence of $\langle\hat \sigma_z\rangle_m$ on $N_{\rm eff}$ is best fitted by a logarithmic function. From this fit, we obtain $N_{\rm eff}\simeq53$ for $d=10$ in the shifted boson Hilbert space. This represents a significant optimization, demonstrating that using only $d=10$ in the shifted space can yield dynamics equivalent to those obtained with $d=53$ in the unshifted space.    

\begin{figure}
  \centering
  \includegraphics[width=1.0\linewidth]{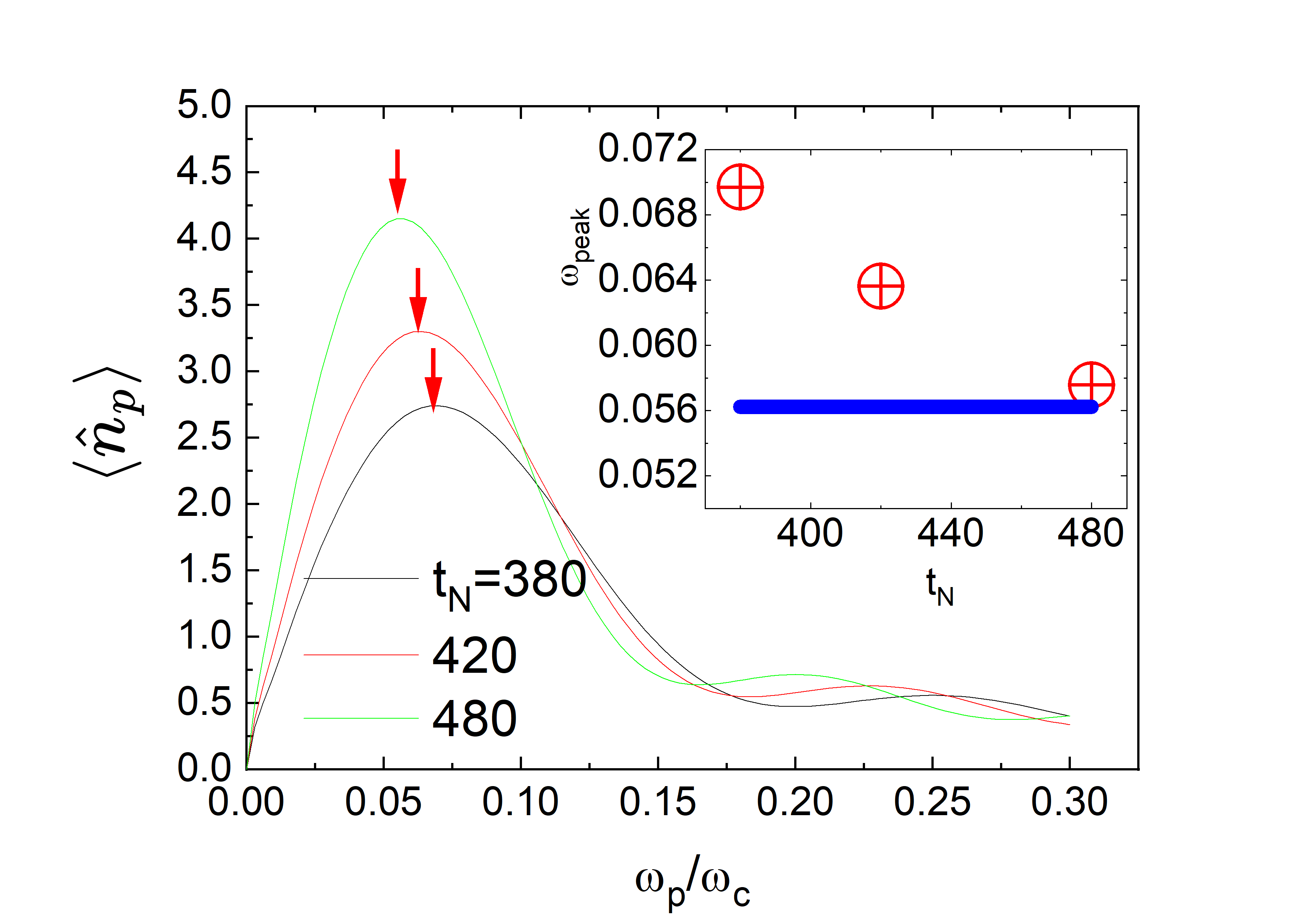}
  \caption{Resonance of the boson modes for the Ohmic SBM at three different time steps ($t_N=320,420,480$) during the TEBD calculation, with inset showing the extracted peak frequency, $\omega_p$, as a function of $t_N$. The other parameters are $s=1.0$, $\alpha=0.1$, $T=0$, $L=30$.}\label{reson}
\end{figure}

To further justify the efficacy and correctness of our method, we also calculate the resonance of a boson bath with an Ohmic spectral density within the shifted boson Hilbert space. We confirm that the resonant frequency of the boson bath remains unaffected by the shift of the boson operators. This resonance reflects an intrinsic property of the heat bath, where boson modes are excited during time evolution, causing energy from the central spin to transfer to the bath. This leads to an accumulation of boson numbers $\langle\hat n_p\rangle$ at the corresponding frequency $\omega_p$, \cite{Chin2016} and a renormalization of the tunneling amplitude $\Delta$ to $\Delta_r$. Here, $\langle\hat n_p\rangle=\langle\hat b^\dagger_p\hat b_p\rangle$ describes the average occupation number of the $p$-th mode in Eq.~\eqref{SBMH}. To obtain $\langle\hat n_p\rangle$, we invert the mapping from Eq.~\eqref{SBMHMap} to Eq.~\eqref{SBMH} to find the occupation in the original $p$-th boson mode. Figure \ref{reson} shows the resonance of the boson modes for the Ohmic SBM at three different time steps $(t_N=320,420,480)$ during the TEBD calculation. As $t_N$ increases, $\omega_p$ converges toward the predicted value of $\Delta(\Delta/\omega_c)^{\alpha/(1-\alpha)}$, \cite{Harris84} as shown in the inset of Fig.~\ref{reson}. The clear convergence of $\Delta_r$ further validates our method.

\begin{figure}
  \centering
  \includegraphics[width=1.0\linewidth]{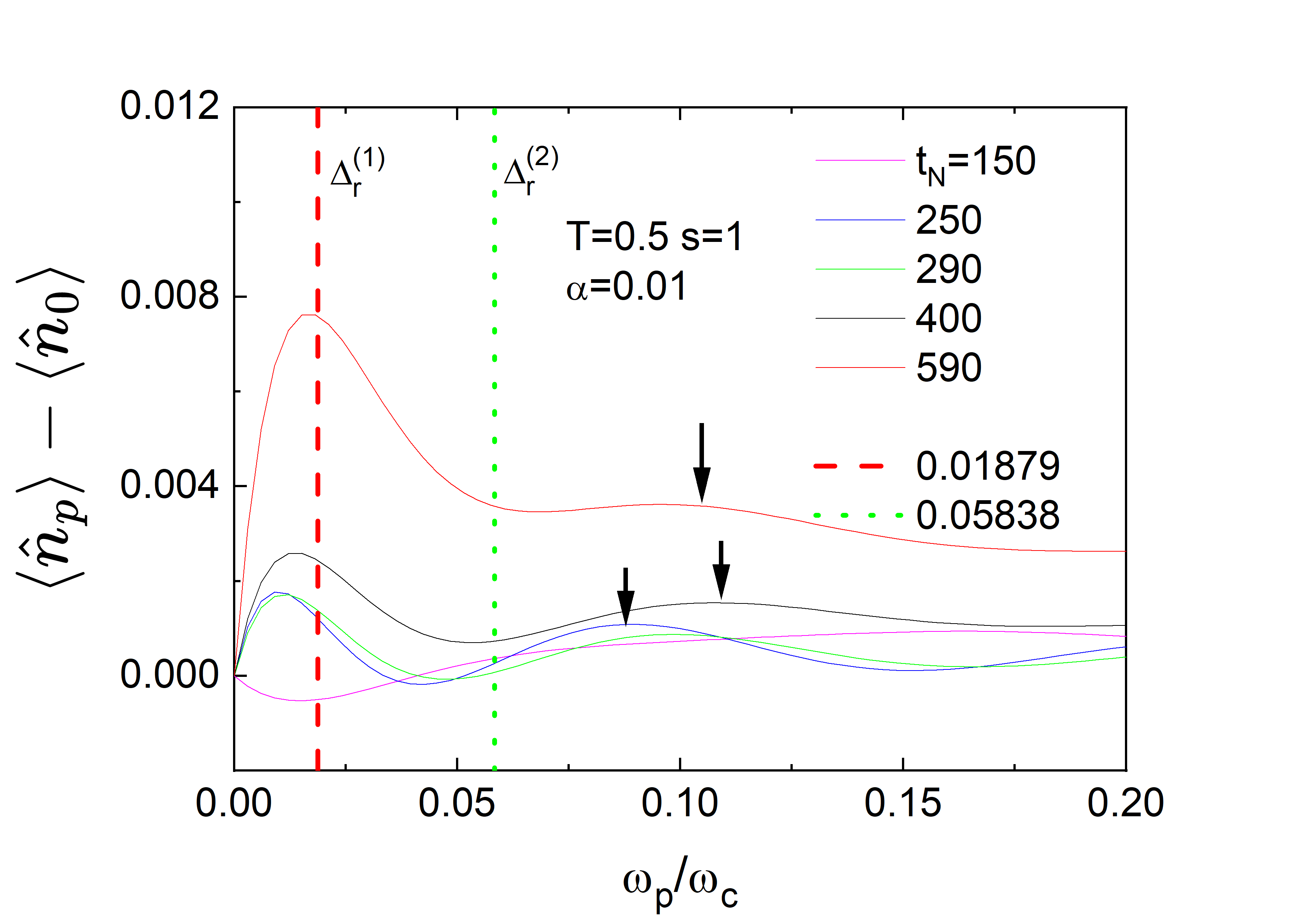}
  \caption{Resonance of the boson modes for five different time steps ($t_N=150,250,290,400,590$) during the TEBD calculation at finite temperature of $T=0.5$. The other parameters are $s=1.0$, $\alpha=0.01$, $L=10$.}\label{fireson}
\end{figure}

We next consider the case of finite temperature, for which the shifted boson basis can also be adapted to simulate the resonance of a polarized boson bath. However, because the trace of the partition function $\rm Tr\exp(-\beta \hat{H})$ must be computed, inserting the shift boson operators $\hat U(\hat x_k)$ into the Trotter-Suzuki decomposition is not beneficial. Consequently, the effective boson number cannot be enhanced at finite temperature using the shifted basis. Despite this, the shifts can still be used to construct the initial density matrix of the polarized boson bath, $\hat{\rho}_B(0)=\exp(-\beta [\hat{H}-\mu\mathcal{E}])/Z_B$, where $\mathcal{E}=\sum_p\lambda_p(\hat{b}_p^{\dagger}+\hat{b}_p)$ and $\mu$ is the polarization parameter. When $\hat H$ is written as a Wilson chain-like Hamiltonian (Eq.~\eqref{SBMHMap}),  $\hat{H}-\mu\mathcal{E}$ can be viewed as the Hamiltonian $\hat{H}'$ in the shifted boson Hilbert space.\cite{LipengChenJCP2023} The shift $\langle\hat x_k\rangle$ is determined by the system-bath coupling $\alpha$ and can be obtained through DMRG calculations, similar to the zero-temperature case.   

The resonance of the boson bath at finite temperature is more complex than that at zero temperature. For the Ohmic SBM, the renormalized tunneling $\Delta_r$ is predicted to obey the self consistent equation \cite{Harris84}
\begin{equation}\label{fdr}
    \frac{\Delta_r}{2}=\frac{\Delta}{2}e^{2\pi\alpha/\beta\Delta_r}[\frac{2+\beta\Delta_r\tanh\beta\Delta_r/2}{\beta\omega_c+\beta\Delta_r\tanh\beta\Delta_r/2}]^\alpha.
\end{equation}
Numerically, Eq.~\eqref{fdr} is found to have two solutions: $\Delta^{(1)}_r$ and $\Delta^{(2)}_r$, with $\Delta^{(1)}_r<\Delta^{(2)}_r<\Delta$. Using TEBD method, we study the resonance for the Ohmic SBM with $\alpha=0.01$ at temperature $T=0.5$, as shown in Fig.~\ref{fireson}. The figure presents the boson occupation number $\langle\hat n_p\rangle$
at different times, with the initial thermal distribution $\langle\hat n_0\rangle$ subtracted. The evolution of $\langle\hat n_p\rangle$ shows that at earlier times, boson modes around $\Delta^{(2)}_r\simeq0.05838$ are excited; however, at later times, the modes finally resonate at $\Delta^{(1)}_r\simeq0.01879$. These results agree well with variational calculations, although the initial resonance at $\Delta^{(2)}_r$ appears to have been neglected in many previous works. It should be noted that for finite-temperature TEBD calculations, the local Fock space of the bosons is doubled ($d^2$) to compute the spin polarization dynamics, and the matrix dimension of the local time evolution operator becomes $d^4\times d^4$. This greatly constrains the feasible value of $d$ (we used $d=4$ in Fig.~\ref{fireson}). Additionally, the shifted basis scheme can introduce significant numerical errors, as detailed in the Supporting Information. 

Although the physical effects of bath polarization on the sub-Ohmic SBM have been extensively investigated, corresponding results for the super-Ohmic case are still scarce. Polarization dynamics in a super-Ohmic environment is often considered coherent over long timescales. However, recent work on the super-Ohmic SBM with a factorized bath initial condition observed pseudo-coherent dynamics at early times and strong system-bath coupling.\cite{NalbachPRA2023} The minimal coupling strength required for this pseudo-coherence dynamics exhibits a non-monotonic dependence on the spectral exponent $s$,\cite{NalbachPRA2023} motivating a thorough study of the quantum dynamics of the super-Ohmic SBM with a shifted bath initial condition. 

\begin{figure}
  \centering
  \includegraphics[width=0.7\linewidth]{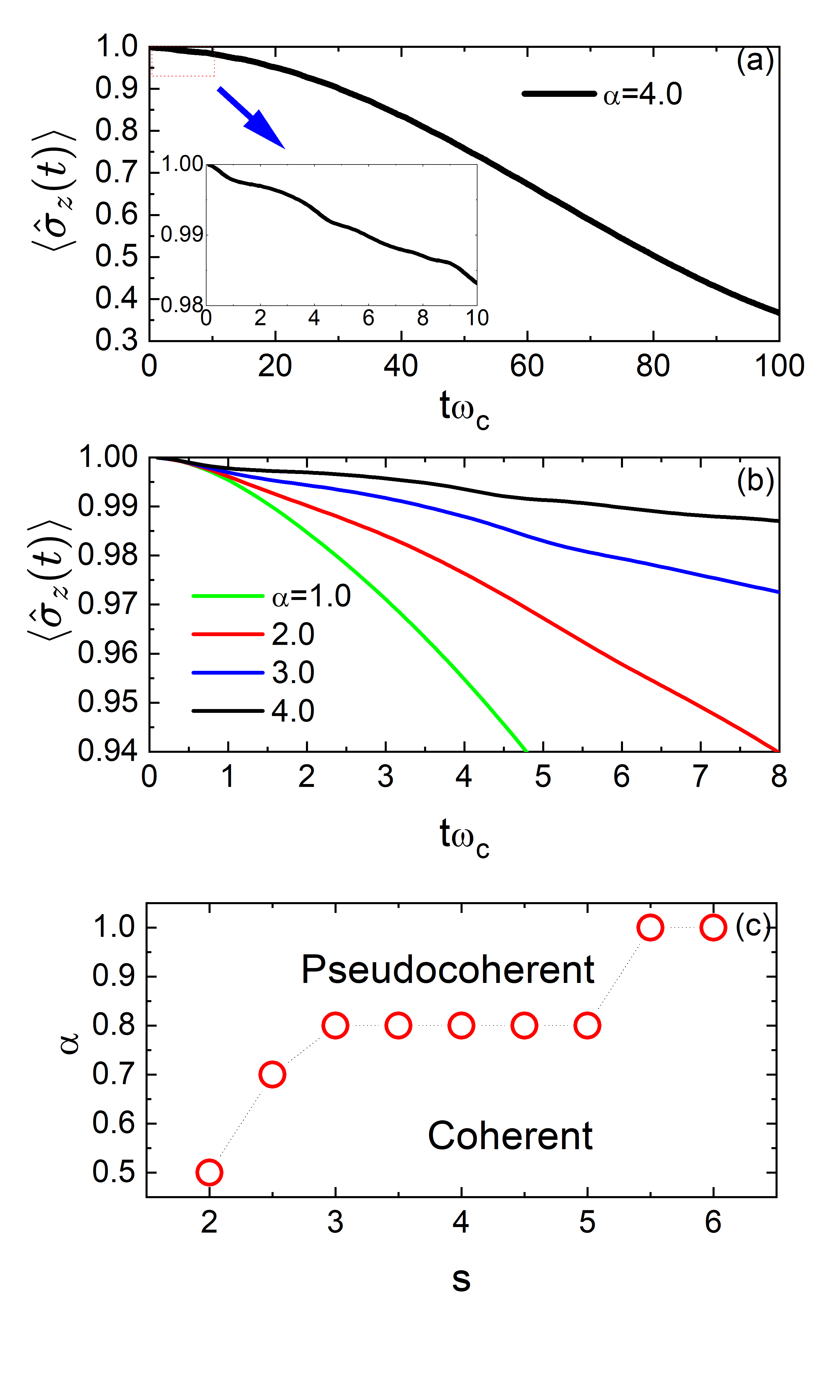}
  \caption{(a) The aperiodic behavior of the spin dynamics at $\alpha=4.0$ for the super-Ohmic SBM, with inset showing the initial part of the curve; (b) Spin dynamics $\langle\hat{\sigma}_z(t)\rangle$ of the super-Ohmic SBM at $\alpha=1.0,2.0,3.0,4.0$; (c) Phase diagram separating pseudocoherent and coherent dynamics for the super-Ohmic SBM. The other parameters are $s=3.0$, $T=0$, $L=30$.}\label{super}
\end{figure}

In this work, we extend our study of polarization dynamics to the super-Ohmic SBM in the shifted boson Hilbert space. By varying the spectral exponent $s$ and the coupling constant $\alpha$, we systematically investigate the dynamics across the parameter space. We identify a new dynamical regime where $\langle\hat \sigma_z(t)\rangle$ exhibits aperiodic pseudo-coherent dynamics that extends over the entire time domain, in remarkable contrast to the behavior observed under a factorized bath initial condition.\cite{supO} Figure \ref{super} (a) shows an example of this aperiodic pseudo-coherent dynamics for $\langle\hat \sigma_z(t)\rangle$ at strong coupling ($\alpha=4$, $s=3$), representing a new type of behavior for the super-Ohmic SBM. For a fixed $s$, this pseudo-coherent dynamics depends on the coupling strength $\alpha$, as shown in Fig.~\ref{super} (b) for $s=3$. Precisely determining the transition point for this dynamical change is challenging due to the lack of a concrete benchmark. As a partial solution, we propose a rough measure based on the oscillatory behavior of $d\langle\hat \sigma_z(t)\rangle/dt$ (see details in Supporting Information). This leads to a schematic phase diagram, shown in Fig.~\ref{super} (c), where the pseudo-coherent dynamical phase emerges above a critical coupling strength $\alpha$. Our numerical results suggest that the polarization dynamics of the super-Ohmic SBM depends strongly on the initial condition of the boson bath and that pseudo-coherent dynamics can be induced by strong system-bath coupling.

In conclusion, we have developed an efficient algorithm by combining the shifted OBB and TEBD methods to study the polarization dynamics of the SBM. Our approach demonstrates a significantly lower computational cost than methods with unshifted boson basis while achieving comparable accuracy, as validated against established zero-temperature results. Furthermore, we have elucidated resonant bath behaviors at both zero and finite temperatures, consistent with variational predictions. Crucially, by applying a shifted bath initial condition to the super-Ohmic SBM, we have identified a new pseudo-coherent dynamics phase. This discovery necessitates a revision of the conventional understanding of coherent dynamics in this regime, highlighting the critical role of initial bath preparation. Our algorithm thus provides an economical and powerful tool for simulating the real-time dynamics of open quantum systems with a large number of boson modes.

\section{Supporting Information}

TEBD method with OBB, numerical errors in OBB, adding boson shifts in DMRG and TEBD, numerical errors for boson shifts, spin dynamics for the super-Ohmic SBM.

\begin{acknowledgement}
Yang Zhao acknowledges support from Hebei Normal University under Grant No. L2023B06 and the Natural Science Foundation of Hebei province, China (Grant No. A2025205002). Lipeng Chen acknowledges support
from the National Natural Science Foundation of China (No. 22473101)
\end{acknowledgement}

\bibliography{SBM}

\end{document}


\section{TEBD method with OBB}
This section details the implementation of the TEBD method with the OBB. The core of the TEBD algorithm involves decomposing the time evolution operator, $\exp(-i\hat Ht)$, into a set of local operators via the Trotter-Suzuki decomposition.\cite{TSD} These local time evolution operators are then applied to an initial MPS. As outlined in Eq.~(3) of the main text, the Hamiltonian is first grouped into even and odd terms: $\hat{H}=\sum_{k\in \mathrm{even}}\hat h_{k,k+1}+\sum_{k'\in \mathrm{odd}}\hat h_{k',k'+1}$. A first-order Trotter-Suzuki decomposition yields 
\begin{align}\label{eq3}
  e^{-i\hat Ht}=& \prod_{M}e^{-i\hat H\Delta t}\tag{S. 1}\\
  =&\prod_{k\in \mathrm{even}}e^{-i\hat h_{k,k+1}\Delta t}\prod_{k'\in \mathrm{odd}}e^{-i\hat h_{k',k'+1}\Delta t}+O(\Delta t^2).\nonumber
\end{align}
Here, $\Delta t = t/M$, $M$ is the number of time steps. The product of local evolution operators can be iteratively applied to the MPS. The error from this decomposition can be reduced by increasing $M$ or by using a higher-order Trotter-Suzuki formula.  

The initial MPS is prepared in the unshifted boson Hilbert space,
\begin{equation}
|\Psi\rangle=\sum_{\sigma=\uparrow,\downarrow}\sum_{\{\vec{n}\}}A^{0}[\sigma]A^{1}[n_1]\cdots{A}^{L-1}[n_{L-1}]|\sigma\rangle|\vec{n}\rangle,\tag{S. 2}
\end{equation}
where  $|\sigma\rangle=|\uparrow\rangle,|\downarrow\rangle$ are the eigenstates of $\hat{\sigma}_z$, and $|\vec{n}\rangle=|n_1,\cdots,n_{L-1}\rangle$
constitutes a basis
of boson-number eigenstates, with  $\hat n_i|n_i\rangle=n_i|n_i\rangle$, $n_i=0,1\cdots d$.
The matrices $A^k[n_k]$ with dimensions ($D_k,D_{k+1}$) contain the MPS variational parameters for the $k$-th site, $L$ is the chain length. Since the bond dimension $D_k$ can grow exponentially with $L$, a maximum truncation dimension $D_c$ is enforced for all $D_k$. Maintaining a large boson truncation $d$ is computationally expensive. To improve efficiency, the OBB method, originally developed for DMRG \cite{OBBDMRG}, is adopted. The OBB is also integrated into the TEBD \cite{TEBD2} and MPS \cite{Chin2016} methods. The core idea of OBB is to replace the $d$ original boson states $|n\rangle$ with a smaller number $d_{\rm opt}<d$ of states $|\tilde{n}\rangle$. These optimized states are the $d_{\rm opt}$ eigenstates with the highest eigenvalues of the single-site reduced density matrix $\rho_{n,n'}[k]=\sum_{D,D'}A^{k*}_{D,D'}[n]A^{k}_{D,D'}[n']$. These eigenstates form a transformation matrix $V^k$ (see Fig.~1 in the main text), which maps the original basis to the optimized basis:$|\tilde n\rangle=\sum_nV^k_{\tilde n,n}|n\rangle$. 

To integrate OBB within the TEBD algorithm, the time evolution operators are first applied to the initial wavefunction $|\Psi(t=0)\rangle$. After applying a local gate $e^{-i\hat h_{k,k+1}\Delta t}$, the two corresponding MPS matrices $A^{k}[n_k]$ and $A^{k+1}[n_{k+1}]$ are updated. The reduced density matrix $\rho_{n,n'}[k]=\sum_{D,D'}A^{k*}_{D,D'}[n]A^{k}_{D,D'}[n']$ for site $k$ is then computed (ensuring $A^{k+1}$ is right-canonicalized first). Diagonalizing this density matrix provides the transformation $V^k$ from its leading $d_{\rm opt}$ eigenvectors. This transformation is applied to project the local basis into the optimized OBB. This process is performed iteratively for each site during the evolution. 
 
\section{Numerical errors in OBB}
\begin{figure}
    \centering
    \includegraphics[width=1.0\linewidth]{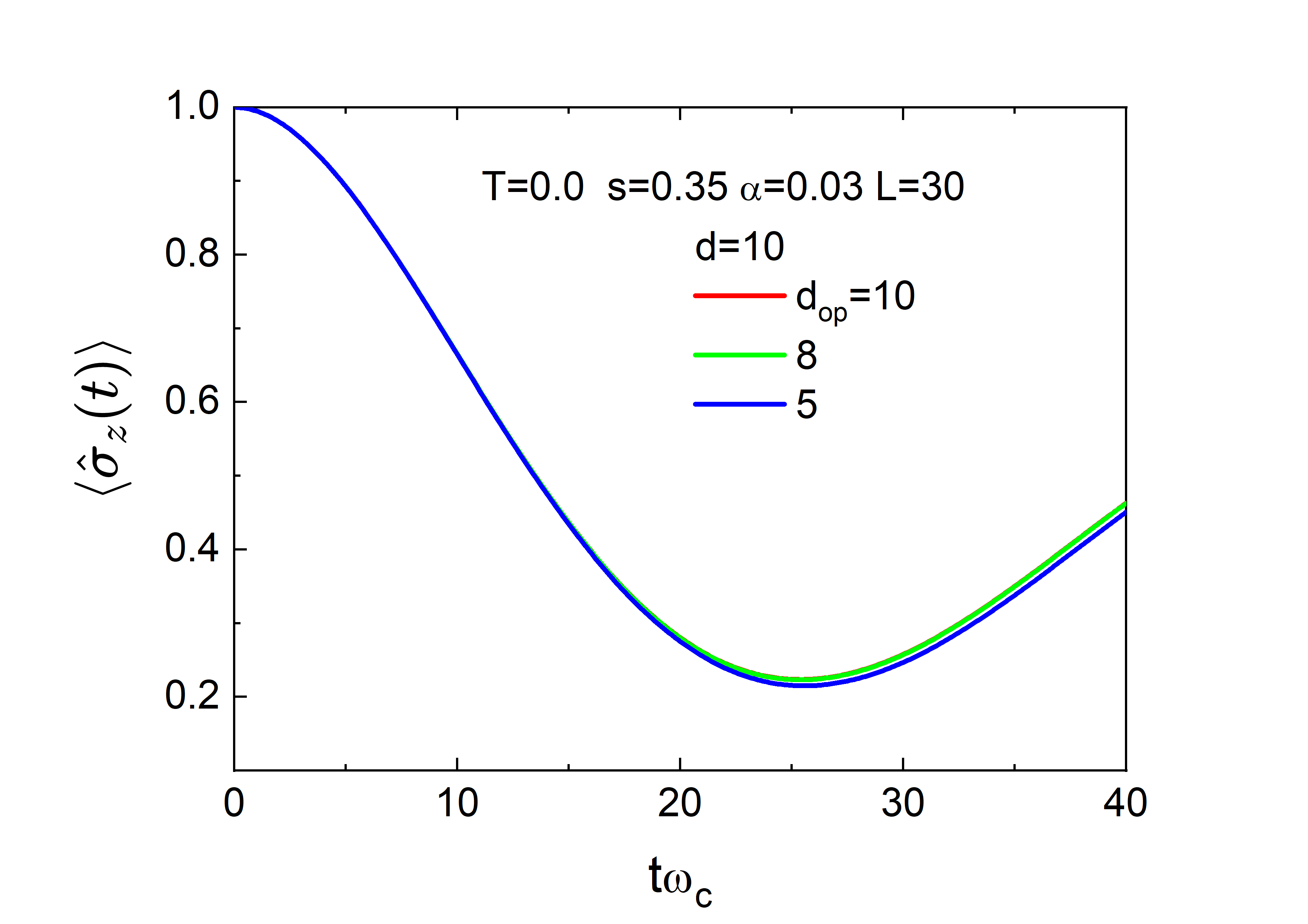}
    \caption{Spin dynamics of  $\langle\hat \sigma_z(t)\rangle$ for different $d_{\rm opt}$ at $d=10$ and zero temperature ($T=0$). The other parameters are $s=0.35$, $\alpha=0.03$, $L=30$.}
    \label{sfig1}
\end{figure}

To construct the OBB, one must select the $d_{\rm opt}$ eigenvectors of the reduced density matrix that correspond to its largest eigenvalues. This section presents our analysis of how the observable $\langle\hat \sigma_z(t)\rangle$ depends on the choice of $d_{\rm opt}$. At zero temperature ($T=0$), as shown in Fig.~\ref{sfig1}, the time evolution of $\langle\hat \sigma_z(t)\rangle$ for a basis size of $d=10$ is virtually identical across different $d_{\rm opt}$ values in the initial time period, even for a modest $d_{\rm opt}=d/2$. Discernible discrepancies emerge only at longer times ($t>20$). This strong agreement is reinforced by truncation errors below $10^{-10}$ for all $d_{\rm opt}$ values, confirming the excellent convergence of the results in Fig.~\ref{sfig1}.  

\begin{figure}
    \centering
    \includegraphics[width=1.0\linewidth]{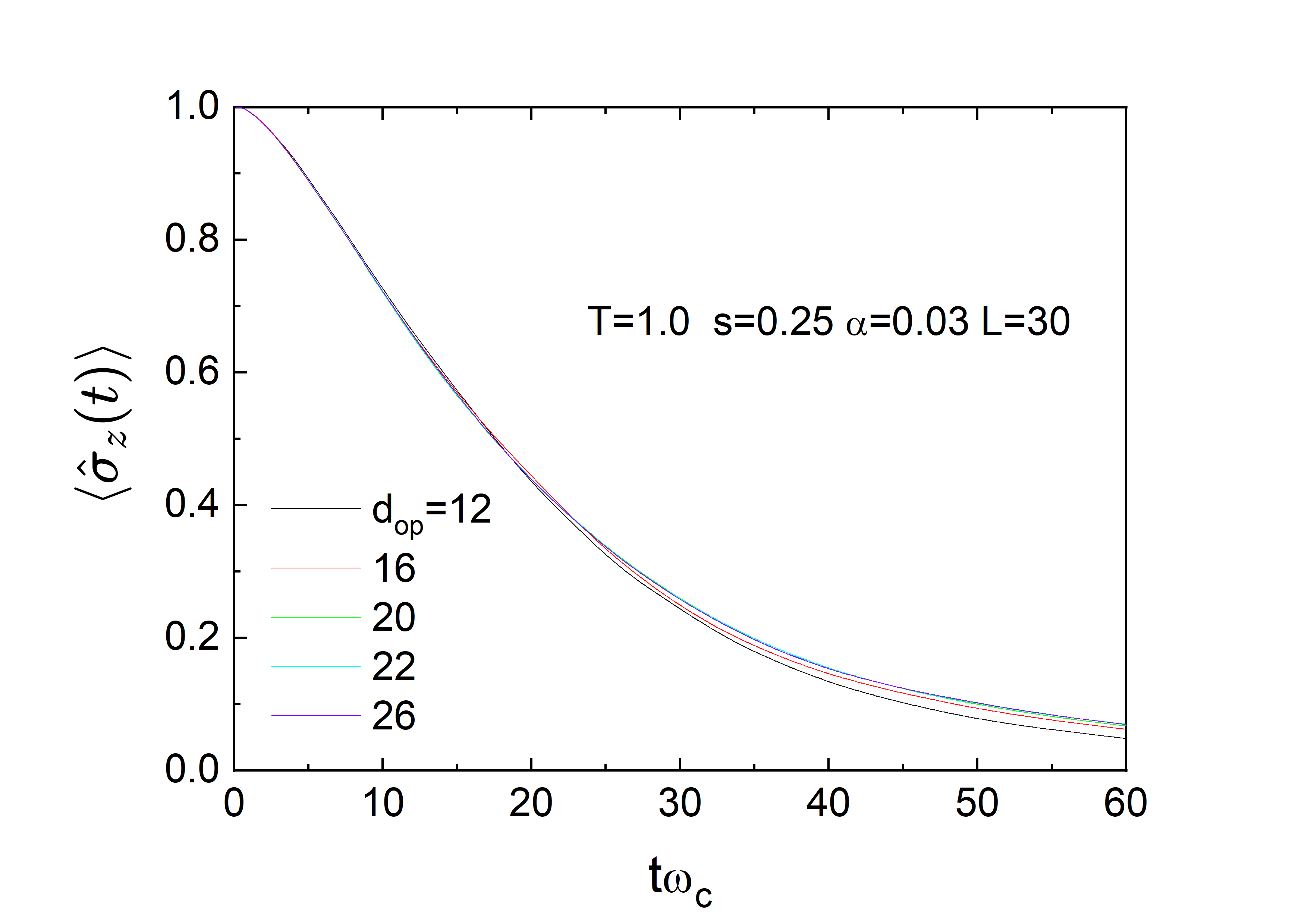}
    \caption{Spin dynamics of  $\langle\hat \sigma_z(t)\rangle$ for different $d_{\rm opt}$ at $d=5$ and finite temperature $T=1.0$. The other parameters are $s=0.25$, $\alpha=0.03$, $L=30$.}
    \label{sfig2}
\end{figure}

\begin{figure}
    \centering
    \includegraphics[width=1.0\linewidth]{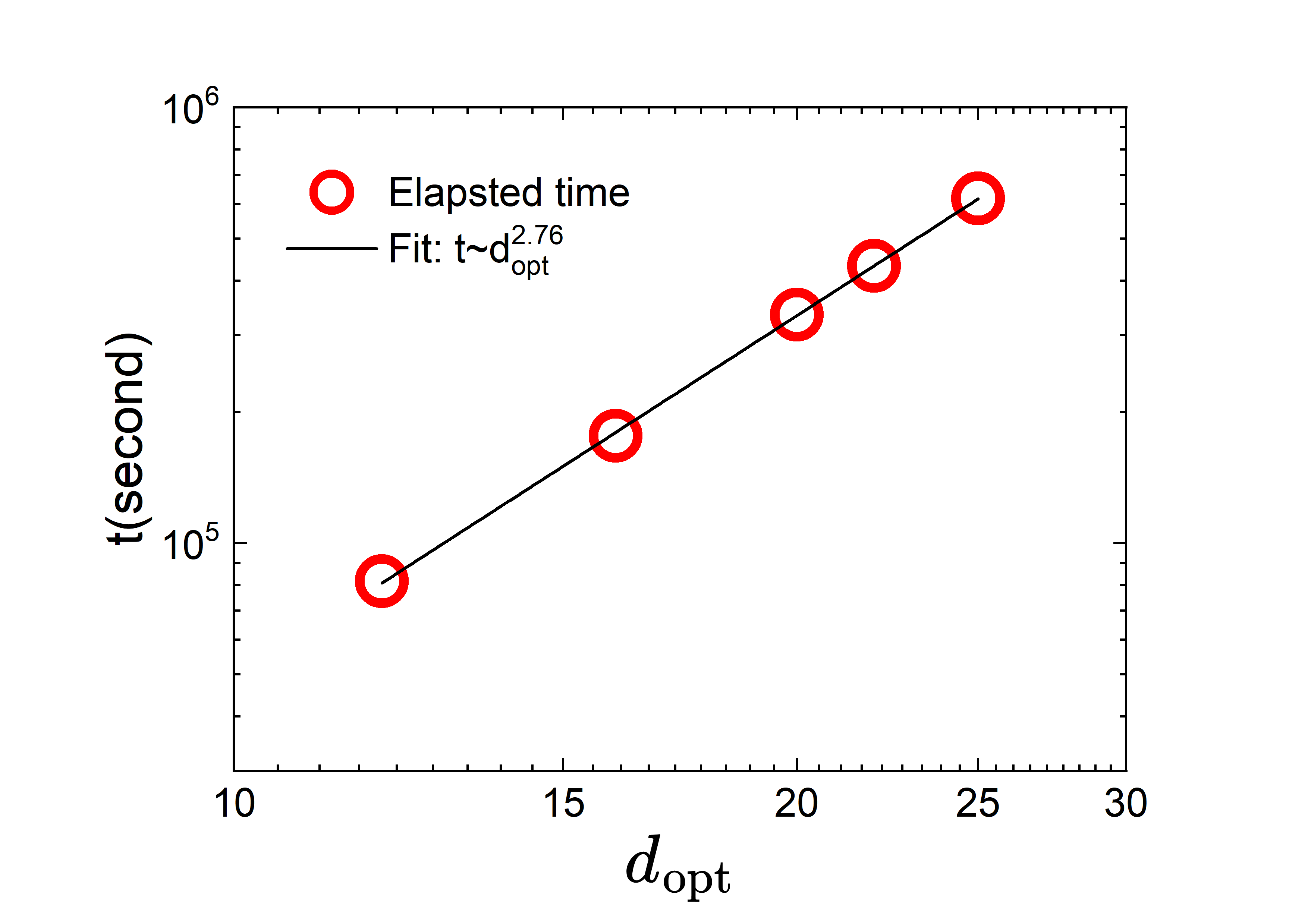}
    \caption{The computational time of the results shown in Fig.~\ref{sfig2} as a function of $d_{\rm opt}$.} 
    \label{sfig3}
\end{figure}

In contrast, at finite temperature (Fig.~\ref{sfig2}), the curves for $\langle\hat \sigma_z(t)\rangle$ at different $d_{\rm opt}$ agree well only within a much shorter initial time window. Furthermore, the truncation error increases rapidly as $d_{\rm opt}$ is reduced. For instance, with $d=5$, the error for $d_{\rm opt}=12$ reaches $10^{-4}$, which is significant. We also find that the computational time scales approximately with $d^{2.76}_{\rm op}$ (Fig.~\ref{sfig3}), rising steeply with larger $d_{\rm opt}$. Therefore, selecting an appropriate $d_{\rm opt}$ at finite temperature requires careful consideration. Based on our empirical findings, setting $d_{\rm opt}=d^2-2$ provides a robust compromise, ensuring accurate results while still yielding substantial savings in computing time. 

\section{Add boson shifts in DMRG}
Although the explicit procedure for adding boson shifts in DMRG is detailed in Ref.~\citenum{guo2}, we provide a concise overview of its key points here. Using boson site $k$ as an example, the local matrix $A^{k}[n_k]$  must be updated during the main DMRG procedure. As noted in the main text, with the use of OBB, $A^{k}[n_k]$ is decomposed into $\sum_{\tilde n_k}\tilde A^{k}[\tilde n_k]V^k_{\tilde n_k,n_k}$. This decomposition requires the sequential updating of both $\tilde A^k[\tilde n_{k}]$ and $V^k_{\tilde n_k,n_k}$ matrices within a single update step. The process follows three steps: (i) update the $\tilde A^k[\tilde n_{k}]$ and $V^k_{\tilde n_k,n_k}$ matrices using standard DMRG algorithms; (ii) calculate the local boson displacement $\langle\hat x_k\rangle$ and shift the local boson operators $\hat b^\dagger_k$ and $\hat b_k$; (iii) return to step (i). This cycle is repeated until $\langle \hat{x}_k \rangle$ converges. The algorithm then proceeds to the next boson site, $k+1$, and repeats the process. Upon completion, this method yields the local boson shifts to all sites and the corresponding wavefunction in the shifted boson Hilbert space.  

\section{Add boson shifts in TEBD}
In contrast to the local update scheme for boson displacement in DMRG, boson shifts can be incorporated into the Hamiltonian using TEBD with a simpler procedure. The basic steps are as follows: (i) apply the time evolution operators to the matrix product state until the energy density converges or another criterion is met; (ii) calculate the boson displacement $\langle\hat{x}_k\rangle$ at each boson site $k$ and shift all local boson operators accordingly; (iii) return to step (i) until the values of $\langle\hat{x}_k\rangle$ meet the chosen convergence standard. Although adding boson shifts in TEBD is methodologically simpler than in DMRG, the calculation of the local expectation values $\langle\hat{x}_k\rangle$ requires significantly more time to achieve convergence. For this reason, we employ the DMRG method to obtain the local boson displacement in the main text. 

\section{Numerical errors for boson shifts}
To shift the boson operators and the initial matrix product state using the operator $\hat U(\hat x_k)=\exp(\varepsilon \langle\hat x_k\rangle(\hat{b}^\dagger_k-\hat{b}_k)/\sqrt{2})$, a moderate value of $\varepsilon$ must be chosen. This is because $\varepsilon=0$ corresponds to no shift, while a large $\varepsilon$ causes $\hat U(\hat x_k)$ to approach a zero operator, as discussed in Ref.~\citenum{guo2}. A critical point is that $\hat U(\hat x_k)$ must first be evaluated in the infinite boson Hilbert space before being truncated. Using the Zassenhaus formula, we expand the boson shift operator as  
\begin{equation}
  U(\hat x_k)=e^{(\langle\hat x_k\rangle(\hat{b}_k^\dagger-\hat{b}_k)/\sqrt{2})}=e^{\langle\hat x_k\rangle\hat{b}_k^\dagger/\sqrt{2}}e^{-\langle\hat x_k\rangle\hat{b}_k/\sqrt{2}}e^{-\frac{\langle\hat x_k\rangle^2}{4}[\hat{b}_k,\hat{b}_k^\dagger]}. \tag{S. 3} 
\end{equation}
The matrix element $\langle n|\hat U(\hat x_k)|m\rangle$ can then be evaluated within the truncated, unshifted boson basis, revealing that $\hat{U}(\hat{x}_k)$ scales approximately as $e^{-\langle\hat x_k\rangle^2}$. 

\begin{figure}
    \centering
    \includegraphics[width=1.0\linewidth]{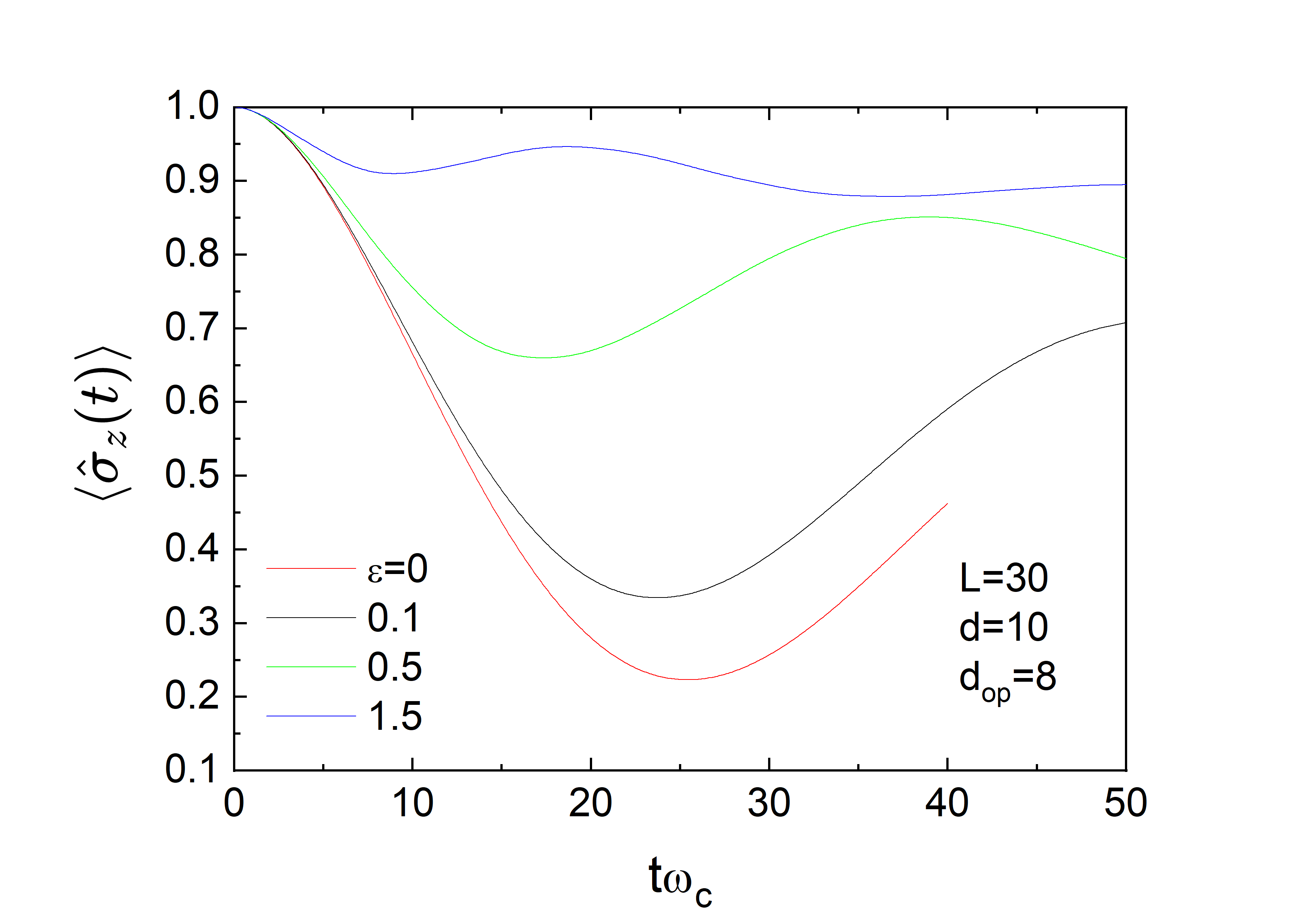}
    \caption{Spin dynamics of $\langle\hat \sigma_z(t)\rangle$ at different $\varepsilon$. The other parameters are $s=0.35$, $\alpha=0.03$, $d=10$, $d_{\rm op}=8$, $L=30$.}
    \label{sfig5}
\end{figure}

Since $\hat U(\hat x_k)$ is constructed in this truncated basis, significant truncation errors arise if $\varepsilon$ is too large. Consequently, simply choosing a large $\varepsilon$ cannot produce a large local effective boson number. To illustrate this, we plot the spin dynamics for different $\varepsilon$ values in Fig.~\ref{sfig5}. The nonphysical oscillations in $\langle\hat \sigma_z(t)\rangle$ for $\varepsilon=1.5$ demonstrate this truncation error. 

Finally, it is important to note that while the distributions of the local boson displacement in the ground state (obtained from various transformations of chain-like Hamiltonian) may differ, this is because the boson operators $\hat{b}_k$ and $\hat{b}^\dagger_k$ are linearly combined through different polynomials (Ref.~\cite{Lmaping}). The physical observables of the SBM, however, remain invariant under these transformations. 
 
\section{Spin dynamics for the super-Ohmic SBM}
\begin{figure}
    \centering
    \includegraphics[width=1.0\linewidth]{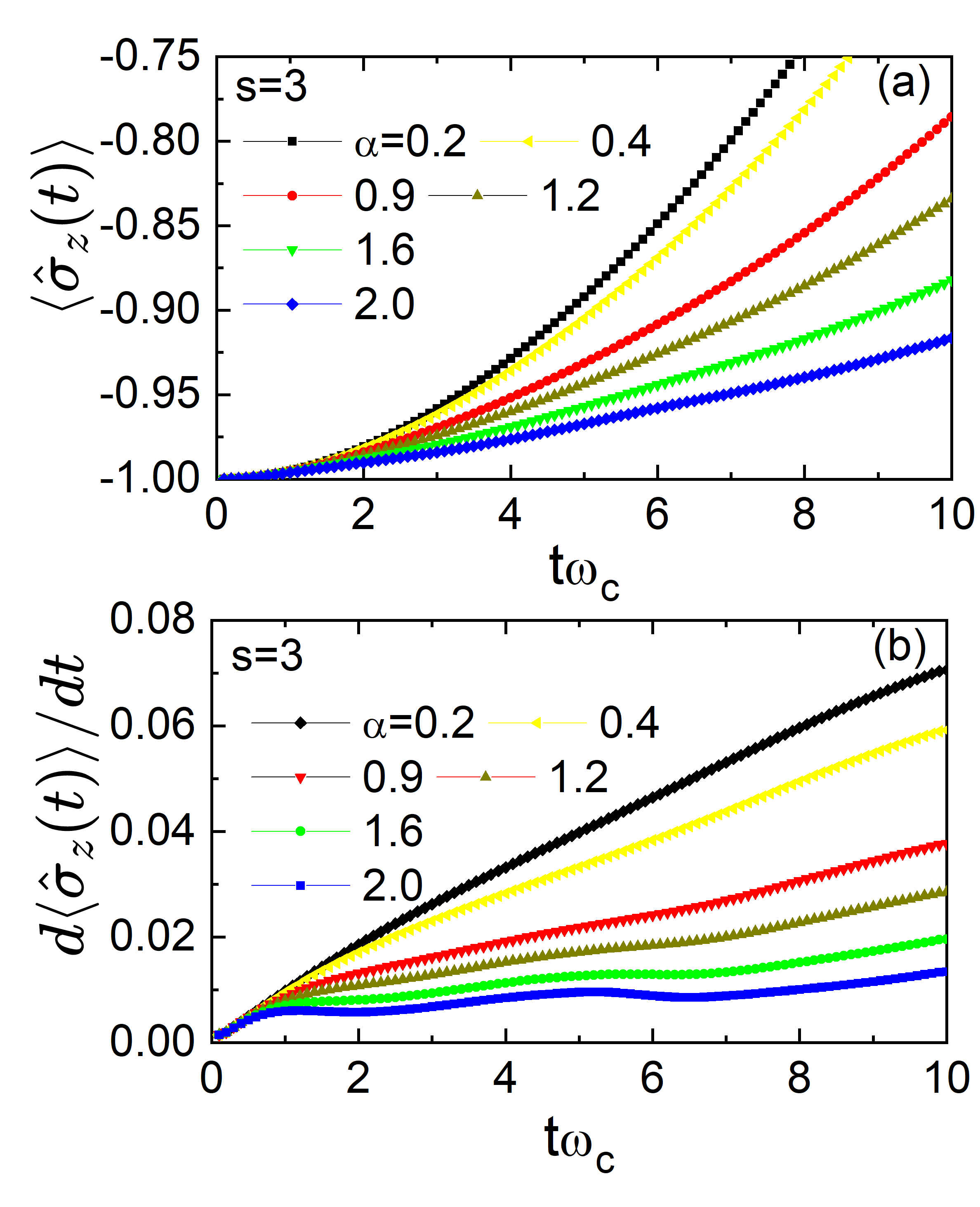}
    \caption{(a) Spin dynamics $\langle\hat \sigma_z(t)\rangle$ and (b) its first derivative with respect to time $d\langle\hat \sigma_z(t)\rangle/dt$ for the super-Ohmic SBM at different $\alpha$ and $s=3.0$.}
    \label{sfig6}
\end{figure}

To map the dynamical phase diagram of the super-Ohmic SBM, we propose a simple criterion: the first derivative of $\langle\hat \sigma_z(t)\rangle$ with respect to time. As shown in Fig.~\ref{sfig6} for the case of $s=3.0$, $d\langle\hat \sigma_z(t)\rangle/dt$ is highly sensitive to changes in the spin polarization. This sensitivity allows it to clearly signal dynamical transitions, thereby enabling us to construct the phase diagram presented in the main text.  

\bibliography{SBM}